\documentclass[reqno,12pt,letterpaper]{amsart}
\usepackage{amsmath,amssymb,amsthm,graphicx,mathrsfs,url,enumerate}
\usepackage[usenames,dvipsnames]{xcolor}
\usepackage[colorlinks=true,linkcolor=Red,citecolor=Green]{hyperref}
\usepackage{amsxtra}
\usepackage{tikz,pgfplots}
\tikzset{invclip/.style={clip,insert path={{[reset cm]
      (-\maxdimen,-\maxdimen) rectangle (\maxdimen,\maxdimen)
    }}}}
\pgfplotsset{%
  every x tick/.style={black, thin},
  every y tick/.style={black, thin},
  every tick label/.append style = {font=\footnotesize},
  every axis label/.append style = {font=\footnotesize},
  compat=1.16
 }

\def\?[#1]{\textbf{[#1]}\marginpar{\Large{\textbf{??}}}}

\let\epsilon=\varepsilon 
\let\phi=\varphi

\newcommand{\D}{{\mathcal D}}
\newcommand{\E}{{\mathcal E}}
\newcommand{\HH}{{\mathcal H}}
\newcommand{\LL}{{\mathcal L}}
\newcommand{\OO}{{\mathcal O}}
\newcommand{\R}{{\mathcal R}}
\newcommand{\RR}{{\mathbb R}}

\newcommand{\ZZ}{{\mathbb Z}}

\newcommand{\UU}{{\mathcal U}}

\newcommand{\NN}{{\mathbb N}}
\newcommand{\CC}{{\mathbb C}}
\newcommand{\CI}{{{\mathcal C}^\infty}}

\newcommand{\CIc}{{{\mathcal C}^\infty_{\rm{c}}}}

\newtheorem{prop}{Proposition}

\newtheorem{hypo}[prop]{Hypothesis}
\numberwithin{equation}{section}
\numberwithin{prop}{section}

\DeclareMathOperator{\diag}{Diag}

\DeclareMathOperator{\Res}{Res}
\DeclareMathOperator{\Spec}{Spec}
\DeclareMathOperator{\comp}{comp}

\DeclareMathOperator{\id}{id}

\let\Im=\Imag
\DeclareMathOperator{\loc}{loc}

\DeclareMathOperator{\rank}{rank}

\DeclareMathOperator{\Span}{span}

\DeclareMathOperator{\supp}{supp}

\DeclareMathOperator{\Diff}{Diff}

\DeclareMathOperator{\Ran}{Ran}

\title{Generic simplicity of resonances in obstacle scattering}
\author{Haoren Xiong}
\email{xiong@math.berkeley.edu}
\address{Department of Mathematics, University of California,
Berkeley, CA 94720, USA}

\begin{document}

\begin{abstract}
    We show that all resonances in Dirichlet obstacle scattering (in $\CC$ in odd dimensions and in the logarithmic cover of $\CC\setminus\{0\}$ in even dimensions) are generically simple in the class of obstacles with $C^k$ (and $C^\infty$) boundaries, $k \geq 2$.
\end{abstract}

\maketitle

\section{Introduction}
\label{introduction}

The evolution of eigenvalues of second order elliptic operators under boundary perturbations have been studied through different perspectives since Hadamard \cite{hadamard}. Uhlenbeck \cite{uhlenbeck} proved generic
properties of eigenvalues and eigenfunctions of second order elliptic operators with
respect to variation of the domain for general boundary conditions. Henry \cite{henry} developed a general theory on perturbation of domains for second order elliptic operators. In this paper we prove that a generic boundary perturbation in obstacle scattering for the Dirichlet Laplacian splits the multiplicities of all resonances in both odd and even dimensions. We formulate the problem as follows:

Suppose that $\OO \subset \RR^n$ is a bounded open set such that $\partial\OO$ is a $C^k$ ($k\geq 2$) hypersurface in $\RR^n$. Let $\Delta_\OO$ be the self-adjoint Dirichlet Laplacian on $\RR^n\setminus \OO$ with domain 
\begin{equation}
\label{eqn:Dirichlet Domain}
    \D(\Delta_\OO) := H^2 (\RR^n\setminus\OO) \cap H_0^1 (\RR^n\setminus\OO).
\end{equation}
The resolvent of $-\Delta_\OO$,
\[  R_\OO(\lambda) := (-\Delta_\OO - \lambda^2)^{-1} : L^2 (\RR^n\setminus\OO) \to L^2 (\RR^n\setminus\OO),\quad\Im\lambda > 0, 
\]
continues meromorphically as an operator from $L_{\comp}^2 (\RR^n\setminus\OO)$ to $L_{\loc}^2 (\RR^n\setminus\OO)$ -- see for instance Dyatlov--Zworski \cite[\S 4.2]{res} and a review in \S\ref{meroCont}. When $n$ is odd the continuation is to $\lambda\in\CC$ and when $n$ is even to the logarithmic cover of $\CC\setminus\{0\}$:
\begin{equation}
\label{eqn:log cover}
\Lambda = \exp^{-1} (\CC\setminus\{0\}).
\end{equation}

We denote the set of poles of $R_\OO(\lambda)$ by $\Res(\OO)$. The elements of $\Res(\OO)$ are called scattering resonances for the obstacle $\OO$. We recall the following facts from \cite[Theorem 4.19]{res} (for $n$ odd) and Christiansen \cite[\S 6]{TJ} (for $n$ even) that:
\begin{equation}
\label{0 isnot resonance}
    0\notin \Res(\OO),\ \textrm{for $n$ odd};\quad 0\textrm{ is not a limit point of $\Res(\OO)$ for $n$ even}.
\end{equation}
Thus for $\lambda \in\Res(\OO)$, its multiplicity $m_\OO(\lambda)$ satisfies
\begin{equation}
\label{eqn:mult obstacle res}
    m_\OO(\lambda) := \rank\oint_\lambda R_\OO(\zeta)d\zeta = \rank\oint_\lambda R_\OO(\zeta)2\zeta d\zeta,
\end{equation}  
where the integral is over a circle containing no other pole of $R_\OO(\zeta)$ than $\lambda$, see \cite[\S 4.2]{res}. A resonance $\lambda\in\Res(\OO)$ is called simple if $m_\OO(\lambda)=1$.

To describe the deformations of obstacles, we follow Pereira \cite{pereira} and introduce a set of $C^k$--smooth mappings ($k\geq 2$) which deforms the obstacle $\OO$:
\begin{equation}
\label{def:Diffeomorphism}
    \Diff(\OO) := \left\{ 
        \begin{gathered}
        \Phi\in C^k(\RR^n;\RR^n) \textrm{ is a $C^k$-diffeomorphism}:\ \Phi(\partial\OO) = \partial\,\Phi(\OO), \\ 
        \textrm{and }\Phi(x)=x,\ \forall\,|x|>R,\quad\textrm{for some }R>0.   
        \end{gathered}
    \right\} .
\end{equation}
Let $X$ be the class of obstacles diffeomorphic to a fixed obstacle $\OO_0$ (for example, $\OO_0=B_{\RR^n}(0,1)$), that is,
\begin{equation}
\label{eqn:X}
    X = \{ \Phi(\OO_0) : \Phi\in\Diff(\OO_0) \} .
\end{equation}  
We introduce a topology in this set by defining a sub-basis of the neighborhoods of a given $\OO\in X$ by
\[  \{ \Phi(\OO) : \Phi\in\Diff(\OO),\ \|\Phi-\id \|_{C^k(\RR^n \setminus\OO)}<\epsilon,\textrm{ with $\epsilon$ sufficiently small.} \}
\]
Now we state the main result of this paper, by a generic set we mean an intersection of open dense sets:

\vspace{0.1in}
\noindent
{\bf Theorem.}
\textit{For any fixed obstacle $\OO_0$ and the corresponding family $X$ given in \eqref{eqn:X}, there exists a generic set $\mathcal{X}\subset X$ such that for every $\OO\in\mathcal X$, all resonances $\lambda\in\Res(\OO)$ are simple.}

\vspace{0.1in}

\noindent
{\bf Remark 1}: We should point out that an analogue of this result for Robin boundary condition (and in particular for the Neumann boundary condition) remains an open problem. The difficulty was overcome by Uhlenbeck \cite{uhlenbeck72} in the case of Neumann eigenvalue problem in a bounded domain $\Omega$ by using Transversality Theorem in infinite dimensions and then deriving a contradiction from the equation $\nabla_{\partial\Omega}u\cdot \nabla_{\partial\Omega}v = \lambda uv$ on $\partial\Omega$ where $\lambda>0$, $u,v\in C^2(\partial\Omega;\RR)$ and $uv\neq 0$ on an open dense subset of $\partial\Omega$, see also \cite[Example 6.4]{henry} for more details. In the case of obstacle scattering with Neumann boundary condition, this argument does not seem to apply for $\nabla_{\partial\Omega}u\cdot \nabla_{\partial\Omega}v = z uv$ when $u,v$ are complex-valued and $z$ is a complex resonance.

\noindent
{\bf Remark 2}: Klopp and Zworski \cite{klopp} proved that a generic potential perturbation in black box scattering (for a definition see for instance \cite[\S 4]{res} and \S \ref{meroCont}) splits the multiplicities of all resonances. This result was extended to scattering on asymptotically
hyperbolic manifolds by Borthwick and Perry \cite{borthwick}, in which the method of complex scaling used in \cite{klopp} was replaced by Agmon's perturbation theory of resonances \cite{agmon}. We will combine the strategies of \cite{klopp} and \cite{borthwick} in the proof of our theorem. However, the boundary perturbation produces additional difficulties.

The paper is organized as follows. In \S\ref{meroCont}
we review the meromorphy of the resolvent $R_\OO(\lambda)$. More precisely, we show that $R_\OO(\lambda)$ admits a meromorphic continuation to $\lambda\in\CC$ in odd dimensions, to $\lambda\in\Lambda$ in even dimensions, as an operator between some weighted Hilbert spaces (instead of $L^2_{\comp} \to L^2_{\loc}$) as an preparation for applying Agmon's perturbation theory of resonances. In \S\ref{sec:agmon} we review Agmon's perturbation theory of resonances \cite{agmon} in which the resonances
are realized as eigenvalues of a non-self-adjoint operator on an abstractly constructed Banach space. We remark that the method of complex scaling is also capable of characterizing resonances in $\CC$ in odd dimensions and resonances in $\Lambda$ in even dimensions with small argument, but Agmon's method allows us to prove the generic simplicity of all resonances in the whole $\Lambda$ in even dimensions. In \S \ref{section: deformation of obstacle} we conjugate the Dirichelet Laplacian of the deformed obstacle $\Phi(\OO)$ by the pullback $\Phi^*$ to obtain an operator on the original domain $\RR^n\setminus\OO$. As a result, the variation of the domain is transferred to the coefficients of the differential operator. In \S\ref{section:agmon theory boundary perturbation}, we use Agmon's theory to study the resonances of the deformed operators introduced in \S\ref{section: deformation of obstacle}. The proof of the theorem is completed in \S \ref{section:proof of generic simplicity} by adapting the strategy in \cite{klopp} to the case of boundary perturbations.

\medskip
\noindent
{\sc Acknowledgments.} The author would like to thank
Maciej Zworski for helpful discussions.
This project was supported in part by
the National Science Foundation grant DMS-1901462.

\section{Meromorphic continuation}
\label{meroCont}
In this section we will follow \cite[\S 4]{res} to introduce a general class of compactly supported self-adjoint perturbations of the Laplacian in $\RR^n$, $P$, which are called black box Hamiltonians.

Let $\HH$ be a complex separable Hilbert space with an orthogonal decomposition:
\begin{equation}
\label{eqn:Hilbert space}
    \HH = \HH_{R_0} \oplus L^2 (\RR^n\setminus B(0,R_0)),
\end{equation}
where $B(x,R)=\{ y\in\RR^n : |x-y|<R \}$ and $R_0$ is fixed. The corresponding orthogonal projections will be denoted by
\[
    u \mapsto u|_{B(0,R_0)}\textrm{ and }u\mapsto u|_{\RR^n\setminus B(0,R_0)},
\]
or simply by the characteristic function $1_L$ of the corresponding set $L$.

We now consider an unbounded self-adjoint operator
\begin{equation}
\label{eqn:P}
    P : \HH \to \HH \quad\textrm{with domain }\D(P).
\end{equation}
Assume that 
\begin{equation}
\label{eqn:domain D_P}
    \D(P)|_{\RR^n\setminus B(0,R_0)} \subset H^2(\RR^n\setminus B(0,R_0)),
\end{equation}
and conversely, $u\in\D(P)$ if $u\in H^2(\RR^n\setminus B(0,R_0))$ and $u$ vanishes near $B(0,R_0)$;
\begin{equation}
\label{eqn:P compact}
    1_{B(0,R_0)} (P + i)^{-1}\textrm{ is compact}.
\end{equation}
We also assume that,
\begin{equation}
\label{eqn:P equals Laplacian}
    1_{\RR^n\setminus B(0,R_0)} Pu = -\Delta (u|_{\RR^n\setminus B(0,R_0)}),\quad\textrm{ for all }u\in\D(P).  
\end{equation}

It is well known that the resolvent $(P-\lambda^2)^{-1}$, $\Im\lambda>0$, $\lambda^2\notin \Spec_\textrm{point}(P)$, 
has a meromorphic continuation to $\CC$ when $n$ is odd; to $\Lambda=\exp^{-1}(\CC\setminus\{0\})$ when $n$ is even: as an operator $(P-\lambda^2)^{-1}:\HH_{\comp}\to \D_{\loc}$, see \cite[Theorem 1.1]{SZ1} and \cite[Theorem 4.4]{res}. However, we need to meromorphically continue $(P-\lambda^2)^{-1}$ as an operator between some Banach spaces to apply Agmon's method \cite{agmon} and prove our theorem. For that we define a weighted subspace of $\HH$ for any large constant $A>0$,
\begin{equation}
\label{eqn:HH_0}
    \HH_0^A := \HH_{R_0}\oplus e^{-A|x|} L^2(\RR^n\setminus B(0,R_0)),
\end{equation}
and a larger space containing $\HH$:
\begin{equation}
\label{eqn:HH_1}
    \HH_1^A := \HH_{R_0}\oplus e^{A|x|} L^2(\RR^n\setminus B(0,R_0)).
\end{equation}
The space $\D_1^A (P)$ is defined using \eqref{eqn:HH_1},
\begin{equation}
\label{eqn:D_1(P)}
    \begin{gathered}
    \D_1^A (P) := \{ u\in\HH_1^A : \chi\in\CIc(\RR^n),\ \chi|_{B(0,R_0)}\equiv 1 \\
    \qquad\qquad\qquad\Rightarrow \chi u\in\D(P),\ \Delta((1-\chi)u)\in\HH_1^A \}.
    \end{gathered}
\end{equation}
We also denote the strips in $\CC$ by
\begin{equation}
\label{eqn:TA}
    T_A := \{  \lambda\in\CC : \Im\lambda > -A \},
\end{equation}
and a family of subsets of $\Lambda$ by 
\begin{equation}
\label{eqn:LambdaA}
\begin{gathered}
    S_m := \{ \lambda \in\Lambda : -m\pi < \arg\lambda < m\pi \},\quad m\in\NN_+, \\
     \Lambda_A := \{ \lambda\in\Lambda: 0<\arg\lambda<\pi \}\cup \{ \lambda\in S_{\lfloor A\rfloor} : |\lambda| < A\}.
\end{gathered}
\end{equation}

We are now ready to state the main result of this section:
\begin{prop}
\label{prop:meromorphic continuation}
Suppose that $P$ is a black box Hamiltonian. Then
\begin{equation}
\label{eqn:resolvent of P}
R(\lambda) := (P-\lambda^2)^{-1} : \HH \to \D(P)\quad\textrm{is meromorphic for }\Im\lambda>0.
\end{equation}
Moreover, when $n$ is odd, the resolvent extends to a meromorphic family
\begin{equation}
\label{eqn:meroCont odd}
    R(\lambda):\HH_0^A \to \D_1^A (P),\quad\lambda\in T_A.
\end{equation}
When $n$ is even \eqref{eqn:meroCont odd} holds with $T_A$ replaced by $\Lambda_A$.
\end{prop}

\noindent
The proof is the same as the one of \cite[Theorem 4.4]{res}. The only difference is that unlike the free resolvent $R_0(\lambda):=(-\Delta-\lambda^2)^{-1}$ mermorphically continued as an operator between $L_{\comp}^2(\RR^n)$ and $H_{\loc}^2(\RR^n)$ there, we have to show that 
\begin{equation}
\label{eqn:free resolvent merocont}
    \begin{gathered}
    \lambda\mapsto R_0(\lambda):e^{-A|x|} L^2(\RR^n) \to e^{A|x|} L^2(\RR^n), \\
    \lambda\mapsto [\Delta,\chi]R_0(\lambda): e^{-A|x|}L^2(\RR^n)\to L_{\comp}^2(\RR^n),\quad\forall\,\chi\in\CIc(\RR^n),
    \end{gathered}
\end{equation}
are meromorphic in $\lambda\in T_A$ when $n$ is odd, $\lambda\in\Lambda_A$ when $n$ is even. 

Denote by $R_0(\lambda,x,y)$ the Schwartz kernel of the free resolvent $R_0(\lambda)$, which can be written in terms of the Hankel functions of the first kind:
\begin{equation}
\label{eqn:Hankel}
    R_0(\lambda,x,y) = c_n \lambda^{n-2} (\lambda|x-y|)^{-\frac{n-2}{2}} H_{\frac{n}{2} - 1}^{(1)}(\lambda|x-y|).
\end{equation} 
We recall some well known facts about $R_0(\lambda,x,y)$ as follows, see for instance \cite[\S 3.1]{res} for a detailed account. When $n$ is odd, \eqref{eqn:Hankel} admits a finite expansion:
\begin{equation}
\label{eqn:R0(x,y) odd}
    R_0(\lambda,x,y) = \lambda^{n-2} e^{i\lambda|x-y|}\sum_{j=\frac{n-1}{2}}^{n-2}\frac{c_{n,j}}{(\lambda|x-y|)^j}.
\end{equation}
For $x\neq y$ this form extends meromorphically to $\lambda\in\CC$.
When $n$ is even, using the relation: 
\begin{equation}
\label{eqn:R0(x,y) even}
    R_0(e^{i\ell\pi}\lambda,x,y) - R_0(\lambda,x,y) =
    c_n \ell (-1)^{\frac{n-2}{2}(\ell+1)} \lambda^{\frac{n-2}{2}}|x-y|^{-\frac{n-2}{2}}J_{\frac{n-2}{2}}(\lambda|x-y|),
\end{equation}
where $J_d(z)$ is the Bessel function, we see that $R_0(\lambda,x,y)$, $x\neq y$ extends to $\lambda\in\Lambda$. In view of \eqref{eqn:R0(x,y) odd}, when $n$ is odd we have the upper bounds for $\lambda\in\CC$:
\begin{equation}
\label{eqn:R0 estimates even}
    |R_0(\lambda,x,y)|\leq 
    \begin{cases}
        C(\lambda)|x-y|^{2-n},\quad & |x-y| \leq |\lambda|^{-1} ; \\
        C(\lambda)e^{-\Im\lambda|x-y|}|\lambda|^{\frac{n-3}{2}} |x-y|^{\frac{1-n}{2}},\quad & |x-y|\geq |\lambda|^{-1}.
    \end{cases}
\end{equation}
For $n$ even, $n\neq 2$, the bounds \eqref{eqn:R0 estimates even} hold for $-\pi<\arg\lambda<2\pi$. This follows from the asymptotics of $H_{d}^{(1)}(z)$, see also Galkowski--Smith \cite{galkowski} for more details. Using \eqref{eqn:R0(x,y) even} and the following formulas about $J_d(z)$:
\[
    J_d(z) \sim \frac{1}{\Gamma(d+1)}\big( \frac{z}{2} \big)^d,\quad \textrm{as }z\to 0,\ \textrm{when }d\in\ZZ,
\]
\[
    J_d(z) = \sqrt{\frac{2}{\pi z}}\left( \cos\big(z-\frac{d\pi}{2}-\frac{\pi}{4}\big) + e^{|\Im z|}\OO\big(|z|^{-1}\big)\right),\quad \textrm{as }|z|\to \infty,\ |\arg z| < \pi.
\]
we can extend \eqref{eqn:R0 estimates even} to any $\lambda\in \Lambda$, $\arg\lambda\leq -\pi$ or $\arg\lambda\geq 2\pi$: 
\begin{equation}
\label{eqn:R0 estimates even new}
    |R_0(\lambda,x,y)|\leq 
    \begin{cases}
        C(\lambda)\,|x-y|^{2-n},\quad & |x-y| \leq |\lambda|^{-1} ; \\
        C(\lambda)\,e^{|\Im\lambda||x-y|}|\lambda|^{\frac{n-3}{2}} |x-y|^{\frac{1-n}{2}},\quad & |x-y|\geq |\lambda|^{-1}.
    \end{cases}
\end{equation}
In the case that $n=2$, $|x-y|^{2-n}$ in \eqref{eqn:R0 estimates even new} is replaced by $-\ln|x-y|$ when $|x-y|\leq |\lambda|^{-1}$.
Now we can conclude from \eqref{eqn:R0 estimates even} and \eqref{eqn:R0 estimates even new} that for any (except possible poles) $\lambda\in T_A$ when $n$ is odd, $\lambda\in\Lambda_A$ when $n$ is even,
\[
    \sup_x \int_{\RR^n} e^{-A|x|-A|y|}|R_0(\lambda,x,y)|dy,\quad  \sup_y \int_{\RR^n} e^{-A|x|-A|y|}|R_0(\lambda,x,y)|dx < \infty.
\]
Using the formula about derivatives of the Hankel functions 
\[
\frac{d}{dz} H_m^{(1)} (z) = H_{m-1}^{(1)}(z) - \frac{m}{z} H_{m}^{(1)}(z),
\]
we can also conclude from the bounds \eqref{eqn:R0 estimates even} that 
\[   
    \sup_x \int_{\RR^n}|[\Delta_x,\chi]R_0(\lambda,x,y)|e^{-A|y|}dy,\quad \sup_y \int_{\RR^n}|[\Delta_x,\chi]R_0(\lambda,x,y)|e^{-A|y|}dx  < \infty.  
\]
Hence \eqref{eqn:free resolvent merocont} follows by the Schur test.

\section{Agmon’s perturbation theory of resonances}
\label{sec:agmon}

In this section we adapt Agmon's perturbation theory of resonances \cite{agmon} to study resonances in obstacle scattering, in which resonances are realized as eigenvalues of a non-self-adjoint operator on an abstractly constructed Banach space. 

Let $\OO\subset\RR^n$ be an obstacle and $\Delta_\OO$ be the corresponding self-adjoint Dirichlet Laplacian on $\RR^n\setminus\OO$, whose resolvent admits a meromorphic continuation by Proposition \ref{prop:meromorphic continuation}. More precisely, for any fixed obstacle $\OO$ and constant $A>0$ let 
\[
B_0 := e^{-A|x|}L^2(\RR^n\setminus\OO),\quad B_1 := e^{A|x|} L^2(\RR^n\setminus\OO),
\]
the resolvent of $-\Delta_\OO$ extends to a meromorphic family 
\[
    (-\Delta_\OO - \lambda^2)^{-1} : B_0 \to \D_1(\OO)\subset B_1,\quad\lambda\in T_A \textrm{ when $n$ odd},\ \lambda\in\Lambda_A\textrm{ when $n$ even},
\]
where $\D_1(\OO)$ is the same as \eqref{eqn:D_1(P)} except that $B_1$ replaces $\HH_1$ there:
\begin{equation}
\label{eqn:D_1 OO}
    \D_1(\OO) = \{ u\in B_1\cap H_{\loc}^2(\RR^n\setminus\OO) : u|_{\partial\OO} = 0,\ \Delta u \in B_1 \},
\end{equation}
and $T_A$, $\Lambda_A$ are given by \eqref{eqn:TA}, \eqref{eqn:LambdaA}. The poles in this meromorphic continuation are called scattering resonances for the obstacle $\OO$.

In view of \eqref{0 isnot resonance}, resonances lie in $T_A\setminus i[0,\infty)$ when $n$ is odd. We consider the map:
\[
    T_A\setminus i[0,\infty)\owns \lambda = r e^{i\theta} \mapsto z = r^2 e^{2i\theta} = \lambda^2 \in\Lambda, 
\]
which is invertible. Throughout this section, we will replace parameter $\lambda$ by $z$ with $z=\lambda^2$ defined above. We introduce the image of $T_A\setminus i[0,\infty)$ or $\Lambda_A$ under this map:
\begin{equation}
\label{defn:D+}
    D_+ := 
    \begin{cases}
        \{ \lambda^2\in \Lambda : \lambda\in T_A\setminus \{0\},\  -\frac{3\pi}{2}<\arg\lambda<\frac{\pi}{2}\} \ &\textrm{ when $n$ is odd}; \\
        \{ z : 0<\arg z<2\pi \}\cup \{ z\in S_{2\lfloor A\rfloor} : |z| < A^2\} \ &\textrm{ when $n$ is even}.
    \end{cases}
\end{equation}
We write the resolvent of $\Delta_\OO$ as follows:
\[
\R(z):=(-\Delta_\OO - z)^{-1} : B_0\to B_1,\quad z\in D_+,
\]
which is a meromorphic family by Proposition \ref{prop:meromorphic continuation}. 
We denote by $\Res(\OO)$, the poles of $\R(z)$, $z\in D_+$, which is also the image of the resonances under the map $\lambda\mapsto z=\lambda^2$.

We note that $-\Delta_\OO$ as an operator acting on $B_1$ is closable. Denote by $P_1$ the closure of $-\Delta_\OO$ in $B_1$, by \eqref{eqn:D_1 OO} we have
\begin{equation}
\label{eqn:P1}
    P_1 = -\Delta : B_1\to B_1\quad\textrm{with domain }\D_1(\OO).
\end{equation}
Let us take $B=L^2(\RR^n\setminus\OO)$, $D=\{ z\in\CC:\Im z >0 \}$. Then one can check that $P:=-\Delta_\OO$ satisfies the abstract hypotheses of Agmon's theory: 
\begin{hypo}
\label{hypo:agmon}
    \begin{enumerate}[(i)]
    \item $P$ is a closed, densely defined operator acting in some Banach space $B$;
    \item The resolvent $(P-z)^{-1}$ is a meromorphic family of operators in $\mathcal{L}(B)$ for $z\in D$;
    \item There are two reflexive Banach spaces $B_0$ and $B_1$ such that $B_0\subset B \subset B_1$;
    \item $P$ as an operator on $B_1$ is closable, and denoting the closure of $P$ in $B_1$ by $P_1$, the resolvent $(P_1 - z)^{-1}$ exists for some $z\in D$ as an operator in $\mathcal{L}(B_1)$;
    \item The resolvent $(P-z)^{-1}$ admits a meromorphic continuation from $D$ to $D_+$ as an operator in $\mathcal{L}(B_0,B_1)$.
\end{enumerate}
\end{hypo}
\noindent
\emph{(iv)} can be seen from the following calculation
\[
   e^{-A|x|} (-\Delta - z) e^{A|x|} = -\Delta - \frac{2Ax}{|x|}\cdot\nabla - \frac{(n-1)A}{|x|} - A^2 - z,
\]
which is invertible as an operator from $\D(\Delta_\OO)$ to $L^2(\RR^n\setminus\OO)$ for $z\in D$, $\Im z\gg A^2$.

Now we fix a resonance $z_0\in\Res(\OO)\subset D_+$, $z_0\neq 0$ then choose $\Sigma$, a bounded domain containing $z_0$, with a $C^1$ boundary $\Gamma$, satisfying
\begin{equation}
\label{condition of Gamma}
    (i)\ \overline{\Sigma}\subset D_+;\quad (ii)\ \Gamma\cap \Res(\OO) = \emptyset;\quad (iii)\ \Sigma \cap D \neq\emptyset.
\end{equation}
Having chosen $\Sigma$ we denote by $B_\Gamma$, the subspace of $B_1$ consisting of elements $f$, admitting a representation of the form:
\begin{equation}
\label{eqn:B_Gamma}
     f = g + \int_\Gamma \R(\zeta)\phi(\zeta)d\zeta,\quad g\in B_0,\ \phi\in C(\Gamma;B_0),
\end{equation}
We recall \cite{agmon} that $B_\Gamma$ is a Banach space with the norm 
\begin{equation}
\label{eqn:B_Gamma norm}
    \|f\|_{B_\Gamma} := \inf_{g,\phi} (\|g\|_{B_0} + \|\phi\|_{C(\Gamma;B_0)})
\end{equation}
where the infimum is taken over all $g\in B_0$ and $\phi\in C(\Gamma;B_0)$ which verify \eqref{eqn:B_Gamma}. Then $B_0\subset B_\Gamma\subset B_1$ are continuous inclusions. Agmon \cite{agmon} also introduced a linear operator $R_\Gamma (z)$ on $B_\Gamma$ associated to any $z\in \Sigma\setminus\Res(\OO)$,
\begin{equation}
\label{eqn:def Resolvent_Gamma}
    R_\Gamma(z) f := \R(z) g + \int_\Gamma (\zeta-z)^{-1} (\R(\zeta)-\R(z))\phi(\zeta)d\zeta,
\end{equation}
where $f\in B_\Gamma$ is given by \eqref{eqn:B_Gamma}. Under Hypothesis \ref{hypo:agmon}, Agmon \cite{agmon} showed that $R_\Gamma(z)$ is a well-defined operator in $\LL(B_\Gamma)$, which is actually the resolvent of an operator $P_\Gamma$ acting on $B_\Gamma$: 
\begin{equation}
\label{eqn:R_Gamma and PGamma}
    R_\Gamma (z) = ( P_\Gamma - z )^{-1}\quad\textrm{for}\quad z\in \Sigma\setminus\Res(\OO),
\end{equation}
where $P_\Gamma$ is closed linear operator in $B_\Gamma$ defined as follows:
\begin{equation}
\label{eqn:PGamma}
    \D(P_\Gamma)=\Ran R_\Gamma(w_0),\quad P_\Gamma u = w_0 u + f
\end{equation}
for $u=R_\Gamma(w_0)f \in \D(P_\Gamma)$, $f\in B_\Gamma$. Here $w_0$ is a fixed point in $\Sigma \cap D$. Moreover, $P_1$ extends $P_\Gamma$ in the sense that
\begin{equation}
\label{eqn:P1 ext PGamma}
    \D(P_\Gamma)\subset \D_1(\OO),\quad P_\Gamma u = P_1 u\quad\textrm{for }u\in\D(P_\Gamma),
\end{equation}
where $\D(P_\Gamma)\subset \D_1(\OO)$ is continuous if they are equipped with the graph norms: 
\[ \|u\|_{\D(P_\Gamma)} := \|u\|_{B_\Gamma} + \|P_\Gamma u\|_{B_\Gamma};\quad \|u\|_{\D_1(\OO)} := \|u\|_{B_1} + \|\Delta u\|_{B_1}. 
\]
We recall from \cite{agmon} the following properties that relate $P_\Gamma$ to $-\Delta_\OO$:
\begin{prop}
\label{prop:Spec PGamma}
$P_\Gamma$ has a discrete spectrum in $\Sigma$, given by $\Res(\OO)\cap \Sigma$. Furthermore, let $z_0\in\Res(\OO)\cap \Sigma$ be an eigenvalue of $P_\Gamma$, $\E_\Gamma(z_0)$ denote the generalized eigenspace of $P_\Gamma$ at $z_0$, then
\begin{equation}
\label{eqn:EGamma}
    \E_\Gamma(z_0) := \left( \oint_{z_0} (P_\Gamma - \zeta)^{-1}d\zeta\right) (B_\Gamma) =  \left(\oint_{z_0} \R(\zeta)d\zeta\right) (B_0) ,
\end{equation}
where the integral is over a circle containing no other resonance than $z_0$. In particular, the multiplicity of $z_0\in\Spec(P_\Gamma)$ satisfies
\begin{equation}
\label{eqn:mult agree}
    m_\Gamma (z_0) := \dim \E_\Gamma(z_0) = m_\OO(\lambda_0),\quad\textrm{with } z_0=\lambda_0^2.
\end{equation}
\end{prop}

Let us now turn to the perturbation theory for resonances. Let $\Omega$ be an open neighborhood of the origin in $\CC$. We assume the following:
\begin{hypo}
\label{hypo:analytic perturbation}
There exists a family of linear
operators $V(t):\D_1(\OO)\to B_0$, $t\in\Omega$, with $V(0)=0$, such that
    \begin{enumerate}
        \item $\|V(t)u\|_{B_0} = \OO(t)\|u\|_{D_1(\OO)},\ \forall\,u\in D_1(\OO)$ as $\Omega\owns t\to 0$;
        \item $\Omega\owns t\mapsto V(t)u$ is an analytic $B_0$-valued function, for any $u\in D_1(\OO)$.
    \end{enumerate}
\end{hypo}
Then we consider a family of operators on $B_\Gamma$:
\begin{equation}
\label{eqn:PGamma t}
    P_\Gamma(t) = P_\Gamma + V(t),\quad\textrm{with domain }\D(P_\Gamma(t)):= \D(P_\Gamma).
\end{equation}
Since $P_\Gamma$ is closed, it follows from the bound in Hypothesis \ref{hypo:analytic perturbation} and a well-known result by Kato \cite{kato2013perturbation} that $P_\Gamma(t)$ is also closed in $B_\Gamma$ provided $|t|$ sufficiently small. Shrinking $\Omega$ if necessary, we assume from now on that $P_\Gamma(t)$ is a closed operator for all $t\in\Omega$, then we can apply analytic perturbation theory to the eigenvalues of $P_\Gamma(t)$ -- see \cite[Chapter VII, \S 1]{kato2013perturbation} for a full treatment.

Fixing a resonance $z_0\in \Res(\OO)\cap \Sigma$, we choose $\Sigma'\Subset \Sigma$ with $z_0 \in \Sigma'$, $z_0$ is also an eigenvalue of $P_\Gamma$ by Proposition \ref{prop:Spec PGamma}. We recall the following perturbation result from \cite[Theorem 7.4]{agmon}:

\begin{prop}
\label{prop:kato perturbation}
There exists an open neighborhood of $0\in\CC$, $\Omega_0\subset\Omega$, such that 
\begin{enumerate}[(i)]
    \item for each $t\in\Omega_0$, $P_\Gamma(t)$ has a discrete spectrum in $\Sigma'$;
    \item the spectrum of $P_\Gamma(t)$ depends analytically on $t$ in the following sense: for each $t\in\Omega_0$, there exist a polynomial $q_t^\Gamma(z)$, of degree independent of $t$, with coefficients analytic in $t$, such that the zeros of $q_t^\Gamma(z)$ in $\Sigma'$ coincide with the eigenvalues of $P_\Gamma(t)$ in $\Sigma'$, with agreement of multiplicities. 
\end{enumerate}
\end{prop}

Shrinking $\Omega_0$ if necessary, we may assume by Hypothesis \ref{hypo:analytic perturbation} that 
\[
    \R(z,t):= (-\Delta_\OO + V(t) - z)^{-1} : B_0\to \D_1(\OO)
\]
exists for $\Im z>c>0$ for all $t\in\Omega_0$. It was shown in \cite[Theorem 7.5]{agmon} that for any $t\in\Omega_0$, $\R(z,t)$ admits a meromorphic continuation with poles of finite rank to $z\in \Sigma'$ given by
\[
    \R(z,t)f = (P_\Gamma(t)-z)^{-1} f,\quad\forall\,f\in B_0,\ z\in \Sigma' \setminus \Spec (P_\Gamma(t)).
\]
The connection between the poles of $z\mapsto\R(z,t)$ and the eigenvalues of $P_\Gamma(t)$ was established in \cite[Theorem 7.7]{agmon}, which shows that these two discrete sets are identical, more precisely, the multiplicity of $z_t$ as an eigenvalue of $P_\Gamma(t)$ equals its rank as a pole of $\R(z,t)$. This correspondence and Proposition \ref{prop:kato perturbation} yield the following perturbation result for resonances -- see \cite[Proposition 8.1]{agmon}:

\begin{prop}
\label{prop:analytic dependence}
Suppose that the multiplicity of resonance $z_0$ equals $m$. Let $K\subset \Sigma'$ be any disc centered at $z_0$ containing no other resonances. Then there exists a neighborhood of $0\in\CC$, $\Omega_0'\subset\Omega_0$, such that for any $t\in\Omega_0'$,
\begin{enumerate}[(i)]
    \item The total rank of the poles of $\R(z,t)$ in $K$ is equal to $m$.
    \item Denote by $z_1(t),\ldots,z_m(t)$ the poles of $\R(z,t)$ in $K$, each repeated with respect to its rank. Then $z_j(t)\to z_0$ as $t\to 0$, $j=1,\ldots,m$.
    \item The average $\hat{z}(t):=m^{-1}\sum_{j=1}^m z_j(t)$ is an analytic function of $t$ in $\Omega_0'$.
\end{enumerate}
\end{prop}

\section{Deformation of obstacle}
\label{section: deformation of obstacle}

In this section we study the deformation of obstacle and the corresponding deformed Dirichlet Laplacian. Let $\OO$ be an obstacle as in \S \ref{introduction}, we use $\Diff(\OO) $ defined in \eqref{def:Diffeomorphism} to describe the deformations of $\OO$. For every $\Phi\in\Diff(\OO)$, $\Phi(\OO)$ is a deformed obstacle satisfying all the requirements as $\OO$ does, thus we can define the Dirichlet Laplacian $\Delta_{\Phi(\OO)}$. We conjugate $\Delta_{\Phi(\OO)}$ by the pullback $\Phi^*$. This will transform the deformed domain $\RR^n\setminus\Phi(\OO)$ to the original one. As a result, the variation is transferred to the coefficients of the newly-defined differential operator. 
For $\OO$, an obstacle, and $\Phi\in\Diff(\OO)$ given in \eqref{def:Diffeomorphism}, the pullback $\Phi^*$ is a bounded operator from $L^2(\RR^n\setminus\Phi(\OO))$ to $L^2(\RR^n\setminus\OO)$, which is invertible with the inverse
$(\Phi^{-1})^*$.
In view of \eqref{eqn:Dirichlet Domain}, the restricted map $\Phi^*: \D(\Delta_{\Phi(\OO)})\to \D(\Delta_\OO)$ is also invertible with the inverse $(\Phi^{-1})^*$, since $\Phi^*$ preserves the Dirichlet boundary condition. Hence we can define the deformed operator $\Delta_\OO^\Phi$ of $\Delta_\OO$ associated to the deformation $\Phi$:
\begin{equation}
\label{eqn:deformed Laplacian}
    \Delta_\OO^\Phi := \Phi^* \Delta_{\Phi(\OO)} (\Phi^{-1})^*: L^2(\RR^n\setminus\OO)\to L^2(\RR^n\setminus\OO),\quad\textrm{with domain }\D(\Delta_\OO).
\end{equation}
Let $J_\Phi^{ij}(x)$ denote $[D\Phi(x)^{-1}]_{ij}$, by a direct calculation we have
\[
    \Phi^* \Delta (\Phi^{-1})^* = \sum_{ij\ell} J_\Phi^{i \ell} J_\Phi^{j\ell}\partial_{x_i x_j}^2 + \sum_{ij\ell m q} (\partial_{x_i x_\ell}^2 \Phi^m) J_\Phi^{j m} J_\Phi^{iq} J_\Phi^{\ell q} \partial_{x_j} ,
\]
where $\Phi^m(x)$ is the $m$-th component of $\Phi(x)=(\Phi^1 (x),\cdots,\Phi^n (x))$.
Now we define
\begin{equation}
\label{eqn:V}
\begin{gathered}
    V := \Delta - \Phi^* \Delta (\Phi^{-1})^* = \sum_{i,j} a_{i j}(x)\partial_{x_i x_j}^2 + \sum_j b_j(x)\partial_{x_j}, \\
    \textrm{where}\quad a_{i j} = \delta_{i j} - \sum_\ell J_\Phi^{i \ell} J_\Phi^{j\ell},\quad b_j =  - \sum_{i \ell m q} (\partial_{x_i x_\ell}^2 \Phi^m) J_\Phi^{j m} J_\Phi^{iq} J_\Phi^{\ell q},
\end{gathered}
\end{equation}
then by \eqref{def:Diffeomorphism} we obtain that for all $1\leq i,j \leq n$,
\begin{equation}
\label{eqn:coeffs V}
    a_{i j}\in C_c^{k-1}(\RR^n),\ b_j \in C_c^{k-2}(\RR^n),\quad \|a_{ij}\|_\infty,\,  \|b_j\|_\infty \leq C\|\Phi - \id\|_{C^2}.
\end{equation}
We note that $-\Delta_{\Phi(\OO)}$ is a self-adjoint black box Hamiltonian, whose resolvent admits a meromorphic continuation by Proposition \ref{prop:meromorphic continuation}. More precisely, 
\[
    ( -\Delta_{\Phi(\OO)} - \lambda^2 )^{-1} : e^{-A|x|} L^2(\RR^n\setminus\Phi(\OO)) \to \D_1(\Phi(\OO)),
\]
is a meromorphic family of operators for $\lambda\in\CC$ when $n$ is odd, $\lambda\in\Lambda$ when $n$ is even. Here $\D_1(\Phi(\OO))$ is defined as in \eqref{eqn:D_1 OO}. Since $\Phi^*$ gives isomorphisms between
\[
    \D_1(\Phi(\OO)) \overset{\Phi^*}{\cong} \D_1(\OO),\quad e^{-A|x|} L^2(\RR^n\setminus\Phi(\OO)) \overset{\Phi^*}{\cong} e^{-A|x|} L^2(\RR^n\setminus\OO)
\]
respectively, it follows from \eqref{eqn:deformed Laplacian} that the resolvent of $-\Delta_\OO^\Phi$ also has a meromorphic continuation given by 
\begin{equation}
\label{eqn:res coincide}
    (-\Delta_\OO^\Phi - \lambda^2)^{-1} = \Phi^* (-\Delta_{\Phi(\OO)}-\lambda^2)^{-1} (\Phi^{-1})^* = \Phi^* R_{\Phi(\OO)}(\lambda) (\Phi^{-1})^*,
\end{equation}
whose poles, denoted by $\Res(-\Delta_\OO^\Phi)$, coincide, with agreement of multiplicities, with the resonances of $\Phi(\OO)$.

\section{Agmon's theory and boundary perturbation}
\label{section:agmon theory boundary perturbation}

In this section we consider Agmon's theory for the deformed operators $-\Delta_\OO^\Phi$. It follows from \eqref{eqn:deformed Laplacian} and \eqref{eqn:V} that $-\Delta_\OO^\Phi$ is also closable on $B_1$ with the closure
\begin{equation}
\label{eqn:P1 Phi}
    P_1^\Phi := \Phi^* (-\Delta) (\Phi^{-1})^* = -\Delta + V : B_1\to B_1\quad\textrm{with domain }\D_1(\OO).
\end{equation}
Thus $-\Delta_\OO^\Phi$ also satisfies the abstract hypotheses of Agmon's theory reviewed in \S\ref{sec:agmon}.

For a fixed domain $\Sigma$ with boundary $\Gamma$ satisfying \eqref{condition of Gamma}, by Proposition \ref{prop:meromorphic continuation} we can assume that
\begin{equation}
\label{eqn:bound resolv on Gamma}
    \sup_{\zeta\in\Gamma} \|\R(\zeta)\|_{\LL(B_0,\D_1(\OO))} < C_\Gamma\quad\textrm{for some constant }C_\Gamma > 0,
\end{equation}
where $\D_1(\OO)$ is equipped with the graph norm:
\[    
    \|u\|_{\D_1(\OO)} = \|u\|_{B_1} + \|\Delta u\|_{B_1}.
\]
Since $V$ defined in \eqref{eqn:V} is a second order differential operator with compactly supported coefficients, it can be viewed as an operator in $\LL(\D_1(\OO),B_0)$ satisfying
\begin{equation}
\label{eqn:norm V}
    \|V\|_{\LL(\D_1(\OO),B_0)} \leq C \|\Phi -\id\|_{C^2},
\end{equation} 
there exists $\delta_\Gamma>0$ sufficiently small such that for all $\zeta\in\Gamma$,
\begin{equation}
\label{eqn:delta_Gamma}
    \|\Phi - \id \|_{C^2}<\delta_\Gamma \implies     \|V\R(\zeta)\|_{\LL(B_0,B_0)} < 1/2,
\end{equation}   
which guarantees that $I+V\R(\zeta): B_0 \to B_0$ is invertible by a Neumann series argument. Thus we have
\begin{equation}
\label{eqn:R_Phi}
    \R_\Phi(\zeta) := (-\Delta_\OO^\Phi - \zeta)^{-1} = \R(\zeta)( I + V\R(\zeta) )^{-1},\quad\zeta\in\Gamma,
\end{equation}
which can be justified first for $\zeta$ near $\Gamma\cap\{z : 0<\Im z<\pi \}$ and then by meromorphic continuation. In particular, $\Gamma\cap \Res(-\Delta_\OO^\Phi)=\emptyset$. Hence for the same domain $\Sigma$ with boundary $\Gamma$, we can define $B_{\Gamma,\Phi}$, $R_{\Gamma,\Phi}$ and $P_{\Gamma,\Phi}$ for the deformed operator $-\Delta_\OO^\Phi$, as in \eqref{eqn:B_Gamma}, \eqref{eqn:def Resolvent_Gamma} and \eqref{eqn:PGamma} with $\R(\zeta)$ replaced by $\R_\Phi(\zeta)$.

Now we explore the relationships between $B_{\Gamma,\Phi}$, $R_{\Gamma,\Phi}$, $P_{\Gamma,\Phi}$ and $B_\Gamma$, $R_\Gamma$, $P_\Gamma$. Assuming that $\|\Phi-\id\|_{C^2}<\delta_\Gamma$, by \eqref{eqn:R_Phi} we have for any $f\in B_\Gamma$,
\[   f = g+\int_\Gamma \R(\zeta)\phi(\zeta)d\zeta = g + \int_\Gamma \R_\Phi(\zeta)(I+V\R(\zeta))\phi(\zeta) d\zeta. 
\]
Since $(I+V\R(\zeta))\phi(\zeta)\in C(\Gamma;B_0)$, $f\in B_{\Gamma,\Phi}$ thus we have $B_\Gamma\subset B_{\Gamma,\Phi}$. Furthermore, \eqref{eqn:delta_Gamma} implies that
\[    
    \|g\|_{B_0} + \|(I+V\R(\zeta))\phi(\zeta)\|_{C(\Gamma;B_0)}
    \leq \frac{3}{2} ( \|g\|_{B_0} + \|\phi\|_{C(\Gamma;B_0)} ),
\]
by taking the infimum as in \eqref{eqn:B_Gamma norm}, we obtain that $\|f\|_{B_{\Gamma,\Phi}}\leq 3/2 \,\|f\|_{B_\Gamma}$. Similarly, for $f\in B_{\Gamma,\Phi}$ we have 
\[   f = g+\int_\Gamma \R_\Phi(\zeta)\phi(\zeta)d\zeta = g + \int_\Gamma \R(\zeta)(I+V\R(\zeta))^{-1}\phi(\zeta) d\zeta \in B_\Gamma, 
\]
and again by \eqref{eqn:delta_Gamma} we can deduce that $\|f\|_{B_\Gamma}\leq 2 \,\|f\|_{B_{\Gamma,\Phi}}$. Therefore,
\begin{equation}
\label{eqn:BGammaPhi}
    B_{\Gamma,\Phi} = B_\Gamma,\ \|\cdot\|_{B_{\Gamma,\Phi}}
    \ \textrm{and }\|\cdot\|_{B_\Gamma}\textrm{ are equivalent},\quad\textrm{if }\|\Phi-\id\|_{C^2}<\delta_\Gamma.
\end{equation}

Henceforth, we identify $B_{\Gamma,\Phi}$ with $B_\Gamma$. Suppose that $f = g + \int_\Gamma \R_\Phi(\zeta)\phi(\zeta) d\zeta \in B_\Gamma$, then for $w_0$ chosen in \eqref{eqn:PGamma}, in view of \eqref{eqn:def Resolvent_Gamma} and \eqref{eqn:R_Phi} we have
\[
    \begin{split}
        R_{\Gamma,\Phi}(w_0) f &= \R_\Phi(w_0)g + \int_\Gamma (\zeta - w_0)^{-1}(\R_\Phi(\zeta) - \R_\Phi(w_0))\phi(\zeta)d\zeta \\
        &= \R(w_0)g_1 + \int_\Gamma (\zeta - w_0)^{-1}(\R(\zeta) - \R(w_0))\phi_1(\zeta)d\zeta,
    \end{split}
\]
where $\phi_1(\zeta) := (I + V\R(\zeta))^{-1}\phi(\zeta) \in C(\Gamma;B_0)$ and 
\[    g_1 := (I + V\R(w_0))^{-1} g + \int_\Gamma \frac{(I + V\R(\zeta))^{-1} - (I + V\R(w_0))^{-1}}{\zeta-w_0}\phi(\zeta)d\zeta \,\in B_0.
\]
Thus $R_{\Gamma,\Phi}(w_0) f = R_\Gamma(w_0) f_1$ for $f_1 := g_1 + \int_\Gamma \R(\zeta)\phi_1(\zeta)d\zeta \in B_\Gamma$, which implies that $\Ran R_{\Gamma,\Phi}(w_0) \subset \Ran R_{\Gamma}(w_0)$. We can also derive that $\Ran R_{\Gamma}(w_0) \subset \Ran R_{\Gamma,\Phi}(w_0)$ by similar arguments. Therefore, recalling \eqref{eqn:PGamma} we obtain that
\[   \D(P_{\Gamma,\Phi}) := \Ran R_{\Gamma,\Phi}(w_0) = \Ran R_{\Gamma}(w_0) = \D(P_\Gamma).
\]
We recall \cite{agmon} that $P_1^\Phi$ extends $P_{\Gamma,\Phi}$ as in \eqref{eqn:P1 ext PGamma}, then for any $u\in\D(P_\Gamma)$, \eqref{eqn:P1 Phi} and \eqref{eqn:P1 ext PGamma} imply that
\begin{equation}
\label{eqn:P1Phi ext PGammaPhi}
    P_{\Gamma,\Phi}u = P_1^\Phi u = P_1 u + Vu = P_\Gamma u + V u
\end{equation}
Hence $P_{\Gamma,\Phi}$ and $P_{\Gamma}$ are related as follows
\begin{equation}
\label{eqn:P_Gamma,Phi}
    P_{\Gamma,\Phi} = P_\Gamma + V : B_\Gamma\to B_\Gamma\quad\textrm{with domain }\D(P_\Gamma). 
\end{equation}
Now we substitute $P_\Gamma$ by $P_{\Gamma,\Phi}$ in Proposition \ref{prop:Spec PGamma} and recall \eqref{eqn:res coincide} to conclude:
\begin{prop}
\label{prop:Spec PGamma+V}
Let $\Sigma$ with boundary $\Gamma$ be chosen as in \eqref{condition of Gamma} and suppose that $\Phi\in\Diff(\OO)$ satisfies $\|\Phi-\id\|_{C^2}<\delta_\Gamma$ for some $\delta_\Gamma>0$ in \eqref{eqn:delta_Gamma}, then $P_{\Gamma,\Phi}$ has a discrete spectrum in $\Sigma$, given by $\Res(\Phi(\OO))\cap \Sigma$.

Furthermore, let $z\in\Res(\Phi(\OO))\cap \Sigma$ be an eigenvalue of $P_{\Gamma,\Phi}$, denote by $\E_{\Gamma,\Phi}(z)$ the generalized eigenspace of $P_{\Gamma,\Phi}$ at $z$, then
\begin{equation}
\label{eqn:EGammaPhi}
    \E_{\Gamma,\Phi}(z) := \left( \oint_{z} (P_{\Gamma,\Phi} - \zeta)^{-1}d\zeta\right) (B_\Gamma) = \left( \oint_{z} \R_\Phi(\zeta) d\zeta\right) (B_0)
\end{equation}
where the integral is over a circle containing no other resonance than $z$. In particular, the multiplicity of $z\in\Spec P_{\Gamma,\Phi}$ satisfies
\begin{equation}
\label{eqn:deformed mult agree}
    m_{\Gamma,\Phi} (z) := \dim \E_{\Gamma,\Phi}(z) = m_{\Phi(\OO)}(\lambda),\quad\textrm{with } z=\lambda^2.
\end{equation} 
\end{prop}

\section{Generic simplicity of resonances in obstacle scattering}
\label{section:proof of generic simplicity}

We will follow the strategy of \cite{klopp} and \cite{borthwick} in the case of potential perturbations to prove our theorem. However we have to overcome the additional difficulties produced by boundary perturbations using the results obtained in \S  \ref{section: deformation of obstacle} and \S \ref{section:agmon theory boundary perturbation}. For simplicity we identify $\CC\setminus i[0,\infty)$ with $\{\lambda\in\Lambda : -3\pi/2 < \arg \lambda < \pi/2\}$ when $n$ is odd.
Let $X$ be the class of obstacles diffeomorphic to a fixed obstacle $\OO_0$ -- see \eqref{eqn:X}, that is for some $k\geq 2$,
\[
     X :=\{\Phi(\OO_0) :
        \Phi\in \Diff(\OO_0)\},
\]
with $\Diff(\OO_0)$ defined by \eqref{def:Diffeomorphism}. We introduce a topology in this set by defining a sub-basis of the neighborhoods of any $\OO\in X$ by
\[
    \mathcal{V}_\epsilon(\OO):=\{\Phi(\OO):\Phi\in \Diff(\OO),\ \|\Phi-\id\|_{C^k} < \epsilon \textrm{ with $\epsilon$ sufficiently small} \}.
\]
For any $\theta_1,\theta_2\in\RR$ and $r>1$, we define
\begin{equation}
\label{eqn:Ethetar}
    \begin{gathered}
    S_{\theta_1,\theta_2}^r := \{ \lambda\in\Lambda : \theta_1< \arg\lambda<\theta_2,\  1/r<|\lambda|<r\}, \\
    E_{\theta_1,\theta_2}^r := \{ \OO\in X : m_\OO(\lambda)\leq 1,\  \forall\,\lambda\in S_{\theta_1,\theta_2}^r \}.
    \end{gathered}
\end{equation}
To prove our theorem it suffices to show that for each $\theta_1,\theta_2$ and $r$, $E_{\theta_1,\theta_2}^r$ is open and dense in $X$, since we can then obtain the generic set $\mathcal{X}$ by taking
\[ \mathcal X := \bigcap_{m=1}^\infty \bigcap_{N=1}^\infty E_{-m\pi,m\pi}^N\ \textrm{when $n$ is even};\quad \mathcal X := \bigcap_{N=1}^\infty E_{-\frac{3\pi}{2},\frac{\pi}{2}}^N\ \textrm{when $n$ is odd}. \]

We proceed the proof of our theorem in steps:

\begin{proof}
Step 1. As in \S \ref{sec:agmon}, for $\OO\in X$ we write $\Res(\OO)$ for the image of resonances under the map $\lambda\mapsto z=\lambda^2$, and for any $z$ the multiplicity is given by $m_\OO(z):=m_\OO(\lambda)$ provided $z=\lambda^2$. Then \[
E_{\theta_1,\theta_2}^r = \{\OO\in X : m_\OO(z)\leq 1,\ \forall\,z\in S_{2\theta_1,2\theta_2}^{r^2} \}.
\]
In view of \eqref{defn:D+}, we can choose $A>0$ large enough, such that $ S_{2\theta_1,2\theta_2}^{r^2} \Subset D_+$ (when $n$ is odd, we only need to check for $\theta_1 = -3\pi/2$, $\theta_2 = \pi/2$). Suppose that there is exactly one resonance $z_0$ in $B(z_0,2\delta)\subset S_{2\theta_1,2\theta_2}^{r^2}$, where $B(z_0,r)$ denotes the disc in $\CC$ centered at $z_0$ with radius $r$. For $\Omega:=B(z_0,\delta)$ we then define
\begin{equation}
\label{eqn:Proj Omega}
    \Pi_\OO(\Omega) := -\frac{1}{2\pi i} \int_{\partial\Omega} (-\Delta_\OO - \zeta)^{-1} d\zeta,\quad m_\OO(\Omega):=\rank\Pi_\OO(\Omega).
\end{equation}
Now we choose a bounded domain $\Sigma$ containing $B(z_0,2\delta)$ with boundary $\Gamma$ satisfying \eqref{condition of Gamma}. We also assume that $\Sigma\Subset S_{2\theta_1,2\theta_2}^{r^2}$. By Proposition \ref{prop:Spec PGamma}, elements in $\Res(\OO)$ coincide with the eigenvalues of $P_\Gamma$ in $\Sigma$. 
In view of \eqref{eqn:mult agree}, we have the relationship:
\begin{equation}
\label{eqn:Proj_Gamma Omega}
    \Pi_\Gamma(\Omega) := -\frac{1}{2\pi i} \int_{\partial\Omega} (P_\Gamma - \zeta)^{-1} d\zeta,\textrm{ then } m_\Gamma(\Omega):=\rank\Pi_\Gamma(\Omega) = m_\OO(\Omega).
\end{equation}
Let $\UU_\epsilon(\OO)$ be a set of deformations defined for small $\epsilon>0$,
\[    \UU_\epsilon(\OO) := \{\Phi\in\Diff(\OO) : \|\Phi-\id\|_{C^k} < \epsilon\}.
\]
Assuming that $\epsilon<\delta_\Gamma$ for constant $\delta_\Gamma$ given in \eqref{eqn:delta_Gamma}, then for every $\Phi\in\UU_\epsilon(\OO)$ Proposition \ref{prop:Spec PGamma+V} implies that
\begin{equation}
\label{eqn:Proj_GammaPhi Omega}
    \Pi_{\Gamma,\Phi}(\Omega) := -\frac{1}{2\pi i} \int_{\partial\Omega} (P_{\Gamma,\Phi} - \zeta)^{-1} d\zeta,\  m_{\Gamma,\Phi}(\Omega):=\rank\Pi_{\Gamma,\Phi}(\Omega) = m_{\Phi(\OO)}(\Omega).
\end{equation}
We recall \eqref{eqn:P_Gamma,Phi} that $P_{\Gamma,\Phi}=P_\Gamma + V$ with $V$ defined in \eqref{eqn:V}, then by \eqref{eqn:norm V} we obtain that if $\epsilon$ is sufficiently small, then for $\zeta\in\partial\Omega$ and $\Phi\in\UU_\epsilon(\OO)$,
\[   (P_{\Gamma,\Phi} - \zeta)^{-1} = (P_\Gamma - \zeta)^{-1} (I + V(P_\Gamma - \zeta)^{-1} )^{-1}
\]
and  $\sup_{\zeta\in\partial\Omega}\|(P_{\Gamma,\Phi} - \zeta)^{-1} - (P_\Gamma -\zeta)^{-1}\|_{B_\Gamma\to B_\Gamma} < C(\Omega)\epsilon $. Then we can derive that $\Pi_\Gamma(\Omega)$ and $\Pi_{\Gamma,\Phi}(\Omega)$ have the same rank for any $\Phi\in\UU_\epsilon(\OO)$ if $\epsilon$ is sufficiently small. We restate this as follows:
\begin{equation}
\label{mult stability}
    m_{\Phi(\OO)}(\Omega) \textrm{ is constant for }\Phi\in\UU_\epsilon(\OO) \textrm{ if $\epsilon$ is sufficiently small}.
\end{equation}
Hence $\OO\in E_\theta^r$ implies that $\{\Phi(\OO):\Phi\in\UU_\epsilon(\OO)\} \subset E_\theta^r$ for some $\epsilon$ sufficiently small, in other words, $E_\theta^r$ is an open subset of $X$. 

\noindent    
Step 2. It remains to show that $E_\theta^r$ is dense in $X$, which is equivalent to:
\begin{equation}
\label{density}
    \forall\,\OO\in X \textrm{ and }\epsilon>0, \quad\exists\,\Phi\in \UU_\epsilon(\OO)\,\textrm{ such that }\Phi(\OO)\in E_\theta^r. 
\end{equation}
Since the number of resonances for the obstacle $\OO$ in $S_{\theta_1,\theta_2}^r$ is finite, it is enough to prove a local statement as it can be applied successively to obtain \eqref{density} (once a resonance is simple it stays simple under small deformations due to \eqref{mult stability}). We  will define $\Omega$ for any given $\OO$ and $z_0\in\Res(\OO)$ as in Step 1, thus to obtain \eqref{density} it suffices to prove that for
\begin{equation}
    \label{Omega density}
    \forall\,\OO\in X,\ z_0\in\Res(\OO) \textrm{ and }\epsilon>0,\quad\exists\,\Phi\in \UU_\epsilon(\OO)\ \textrm{ s.t. }m_{\Phi(\OO)}(z)\leq 1,\ \forall\,z\in\Omega.
\end{equation}
To establish \eqref{Omega density} we proceed by induction. We note that for each $\OO\in X$, $z_0\in\Res(\OO)$, one of the following cases has to occur:
\begin{equation}
\label{good case}
    \forall\,\epsilon>0,\quad\exists\,\Phi\in\UU_\epsilon(\OO)\ \textrm{ s.t. }1\leq m_{\Phi(\OO)}(z)< m_{\Phi(\OO)}(\Omega),\quad \forall\,z\in\Omega,
    \end{equation}
or
\begin{equation}
\label{bad case}
    \exists\,\epsilon>0,\ \textrm{s.t.}\quad\forall\,\Phi\in\UU_\epsilon(\OO),\ \exists
    z=z(\Phi)\in\Omega,\ m_{\Phi(\OO)}(z) = m_{\Phi(\OO)}(\Omega)>1.
\end{equation}
The first possibility means that by applying an arbitrarily small deformation $\Phi$ to $\OO$ we can obtain at least two distinct resonances for $\Phi(\OO)$ in $\Omega$. The second
possibility means that under any small deformations the maximal multiplicity persists.

\noindent
Step 3. Assuming \eqref{good case} we can prove \eqref{Omega density} by induction on $m_\OO(z_0)$. If $m_\OO(z_0)=1$ there is nothing to prove. Assuming that we proved \eqref{Omega density} in the case $m_\OO(z_0) < M$, we now assume that $m_\OO(z_0)=M$. We note that for any $\Phi_1\in\Diff(\OO)$ and $\Phi_2\in\Diff(\Phi_1(\OO))$, there exists $C=C(k,n)$ such that
\[
\|\Phi_2\circ\Phi_1 - \id\|_{C^k} \leq C \big(\|\Phi_2 - \id\|_{C^k} + \|\Phi_1 - \id\|_{C^k}\big).
\]
In view of \eqref{good case} we can find $\Phi_0\in\Diff(\OO)$ with $\|\Phi_0 - \id\|_{C^k} < \epsilon/(2C)^M$ such that $m_{\Phi_0(\OO)}(\Omega)=m_\OO(\Omega)$ (using \eqref{mult stability}) and that all resonances in $\Omega$, denoted by $z_1,\cdots,z_\ell$, satisfy $m_{\Phi_0(\OO)}(z_j)<M$. We now find $r_j$ such that 
\[ B(z_j,2r_j)\subset\Omega,\ \{z_j\}=B(z_j,2r_j)\cap\Res(\Phi_0(\OO)),\quad B(z_j,2r_j)\cap B(z_i,2r_i) = \emptyset. \]
We put $\Omega_j:= B(z_j,r_j)$ and apply \eqref{Omega density} successively to $\Phi_{j-1}\circ\cdots\circ\Phi_0(\OO)$, $j=1,\ldots,\ell$ with $\|\Phi_j - \id\|_{C^k} < \epsilon/(2C)^{\ell +1-j}$ (by \eqref{mult stability} we can assume that $\Phi_j$ is sufficiently close to the identity map such that resonances in $\Omega_0,\cdots,\Omega_{j-1}$ that are already simple stay simple while total multiplicities in $\Omega_{j+1},\cdots,\Omega_\ell$ are invariant). Then we obtain the desired $\Phi = \Phi_\ell \circ\cdots\circ\Phi_0 \in \UU_\epsilon(\OO)$ since (note that $\ell<M$)
\[ \|\Phi_\ell\circ\cdots\circ\Phi_0 -\id\|_{C^k} < \sum_{j=1}^\ell C^{\ell+1-j} \frac{\epsilon}{(2C)^{\ell+1-j}} + C^\ell \frac{\epsilon}{(2C)^M} \leq \epsilon. \]

\noindent
Step 4. It remains to show \eqref{bad case} is impossible. For that, we shall argue by contradiction. Suppose that there exist an obstacle $\OO\in X$ and a resonance $z_0\in \Omega$ with some disc $\Omega=B(z_0,r)$ containing no other resonances, such that \eqref{bad case} holds. In fact we may assume further that $\OO$ has $\CI$-boundary since we can deform $\OO$ to a smooth obstacle $\Tilde{\OO}$ through some $\Tilde{\Phi}\in\Diff(\OO)$ with $\|\Tilde{\Phi}-\id\|_{C^k}\ll \epsilon$, decreasing $\epsilon$ if necessary, then \eqref{bad case} still holds with $\Tilde{\OO}$ and $\Tilde{z}_0 = z(\Tilde{\Phi})$ replacing $\OO$ and $z_0$. Hence we assume in the following that $\OO$ is a smooth obstacle.

Let $M = m_{\OO}(\Omega)$. Suppose that $\Sigma$ and $\Gamma$ are chosen as in Step 1. Using \eqref{eqn:Proj_Gamma Omega} and \eqref{eqn:Proj_GammaPhi Omega} we obtain an equivalent statement to \eqref{bad case}:
\begin{equation}
\label{bad case 1}
        \exists\,\epsilon>0,\ \textrm{s.t.}\quad\forall\,\Phi\in\UU_\epsilon(\OO),\ \exists\,
    z=z(\Phi)\in\Omega,\ m_{\Gamma,\Phi}(z) = m_{\Gamma,\Phi}(\Omega) > 1.
\end{equation}
For $\Phi\in\UU_\epsilon(\OO)$, we define 
\[    q(\Phi) := \min\{ q \in \mathbb{N} : (P_{\Gamma,\Phi} - z(\Phi))^q \Pi_{\Gamma,\Phi}(\Omega) = 0 \},
\]
then $1\leq q(\Phi)\leq M$. It follows from \eqref{eqn:P_Gamma,Phi} and \eqref{eqn:V} that if $\|\Phi_j-\Phi\|_{C^{2M}}\to 0$ and $ (P_{\Gamma,\Phi_j} - z(\Phi_j))^q \Pi_{\Gamma,\Phi_j}(\Omega) = 0 $, then $ (P_{\Gamma,\Phi} - z(\Phi))^q \Pi_{\Gamma,\Phi}(\Omega) = 0 $. We now define 
\[ q_0 := \max\{ q(\Phi) : \Phi\in \UU_{\epsilon/2}(\OO) \}, \]
and assume that the maximum is attained at $\Phi_0$ i.e. $q(\Phi_0)=q_0$, then there exists $\epsilon'>0$ such that \[
\|\Phi-\Phi_0\|_{C^{2M}} < \epsilon' \implies q(\Phi)=q_0.
\]
Therefore, we can choose a $\Tilde{\Phi}_0\in\Diff(\OO)$ that is in $\CI(\RR^n;\RR^n)$ with $\| \Tilde{\Phi}_0 - \Phi_0\| \ll \epsilon'$. Replacing $\OO$ in \eqref{bad case 1} by $\Tilde{\Phi}_0(\OO)$ and decreasing $\epsilon$ such that $\epsilon\ll \epsilon'$, we assume in the following that
\begin{equation}
\label{bad case 2}
    \begin{gathered}
    \forall\,\Phi\in\Diff(\OO),\ \|\Phi-\id\|_{C^{2M}} < \epsilon,\ \exists\,z(\Phi)\textrm{ and }1\leq q_0\leq M \textrm{ such that}\\ m_{\Gamma,\Phi}(z(\Phi)) = \rank \Pi_{\Gamma,\Phi}(\Omega) = M > 1, \\
    ( P_{\Gamma,\Phi} - z(\Phi) )^{q_0} \Pi_{\Gamma,\Phi}(\Omega) = 0,\quad ( P_{\Gamma,\Phi} - z(\Phi) )^{q_0 - 1} \Pi_{\Gamma,\Phi}(\Omega) \neq 0. \end{gathered}
\end{equation}

\begin{figure}[h]
    \centering
    \begin{tikzpicture}[scale=1]

\draw [line width=0.7pt] plot [smooth cycle, tension=1] coordinates {(-6.6,2.7) (-6.4,1) (-4.5,0.4) (-3.5,1.5) (-3.3,3)  (-4,4.7) (-5.5,4.5)};
\begin{scope}
\clip (3,3.5) rectangle (4,2);
\draw [line width=0.7pt] [dashed] plot [smooth cycle, tension=1] coordinates {(0.4,2.7) (0.6,1) (2.5,0.4) (3.5,1.5) (3.7,3)  (3,4.7) (1.5,4.5)};
\end{scope}

\begin{scope}
\begin{pgfinterruptboundingbox}
\path[invclip] (3,3.5) rectangle (4,2);
\end{pgfinterruptboundingbox}

\draw [line width=0.7pt] plot [smooth cycle, tension=1] coordinates {(0.4,2.7) (0.6,1) (2.5,0.4) (3.5,1.5) (3.7,3)  (3,4.7) (1.5,4.5)};
\end{scope}

\draw [line width=0.7pt] plot [smooth] coordinates { (3.64,3.43) (3.63,3.25) (3.57,3) (3.45,2.65) (3.6, 2.4) (3.74,2.2) (3.73,2.01) };

\path[->] [line width=0.7pt] (-2.88,2.5) edge [bend left] node[above] {$\varphi_t^h$} (-0.1,2.5);

\node at (-3.26,2.72) {$\bullet$};

\path[->] (-3.26,2.72) edge node[above] {$V_h$} (-4.2,2.6);

\node at (-3,3) {$x_0$};

\node at (-4.8,3.2) {$\OO$};

\node at (2.2,3.2) {$\phi_t^h(\OO)$};


\end{tikzpicture}
    \caption{Deformation $\phi_t^h$ in $\Diff(\OO)$ acting near a fixed point on $\partial\OO$, which is used in Step 5 of the proof.}
    \label{fig:deformation}
\end{figure}
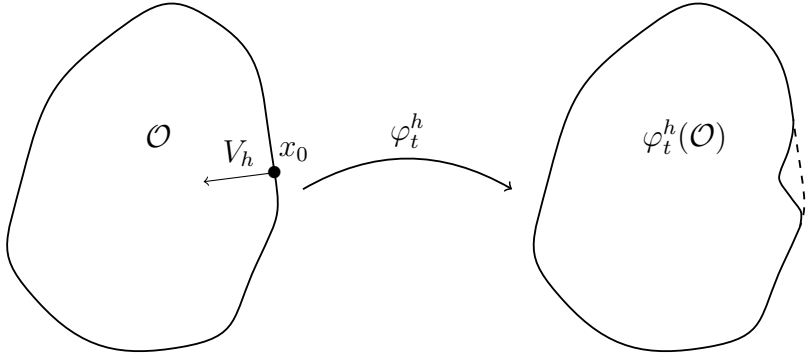

\noindent
Step 5. Before proving \eqref{bad case 2} is impossible we introduce a family of deformations in $\Diff(\OO)$ acting near a point on $\partial\OO$. For any fixed $x_0\in\partial\OO$, we consider the normal coordinates near $x_0$, that is there is some $U=B_{\RR^n}(x_0,2r_0)$ such that for each $x\in U$ there exist unique $(x',x_n)\in \partial\OO\times\RR$ with $x = x' + x_n \nu(x')$, where $\nu(x')$ is the normal vector at $x'$ pointing to the interior of $\OO$. Let $\rho\in\CIc(\RR;[0,1])$ be a bump function such that $\rho(0)=1$ and $\supp\rho\subset(-r_0,r_0)$. Fixing $h_0>0$ small, we choose a family of functions $\chi_h\in \CI(\partial\OO;[0,\infty))$ depending continuously in $h\in(0,h_0]$ such that
\begin{equation}
\label{eqn:chi_h}
    \int_{\partial\OO} \chi_h(x') dS(x') = 1,\quad \supp\chi_h\subset B_{\partial\OO}(x_0,h)\subset U,\quad\forall\,h\in(0,h_0],
\end{equation}
where $B_{\partial\OO}(x_0,h)$ is a geodesic ball on $\partial\OO$ with center $x_0$ and radius $h$. For each $h\in (0,h_0]$, we construct a smooth vector field $V_h\in\CIc(\RR^n;\RR^n)$ as follows
\begin{equation}
\label{eqn:Vh}
    \begin{gathered}
        V_h(x) = \chi_h(x') \rho(x_n) \nu(x'),\ \textrm{for } x = x'+x_n\nu(x') \in U, \\ 
        \textrm{and }V_h(x) = 0\textrm{ for all }x\in\RR^n\setminus U.
    \end{gathered}    
\end{equation}
Then we introduce a family of smooth deformations produced by $V_h$:
\begin{equation}
\label{phi h deformation}
    \varphi_t^h \in \CI(\RR^n;\RR^n),\quad \varphi_t^h (x) := x + t V_h(x).
\end{equation}
It follows from \eqref{eqn:Vh} that for every $h\in(0,h_0]$ there is $t_0=t_0(h)\ll 1$ such that
\[    \forall\,t\in(-t_0,t_0),\quad\varphi_t^h\in\Diff(\OO),\ \|\varphi_t^h-\id\|_{C^{2M}} < \epsilon.
\]

\noindent
Step 6. To show that \eqref{bad case 2} is impossible we first assume the case $q_0>1$. We recall \eqref{eqn:EGamma} that $\Pi_{\Gamma}(\Omega)(B_\Gamma) = \Pi_{\Gamma}(\Omega)(B_0)$, let \[
\CIc(\RR^n\setminus\overline{\OO}) := \{f\in\CIc(\RR^n) : \supp f\subset \RR^n\setminus\overline{\OO} \}
\]
then $\Ran\Pi_{\Gamma}(\Omega) = \Pi_{\Gamma}(\Omega)(\CIc(\RR^n\setminus\overline{\OO}))$ since $\Pi_{\Gamma}(\Omega)$ is finite rank and $\CIc(\RR^n\setminus\overline{\OO})$ is dense in $B_0$. Thus by \eqref{bad case 2} we can find $w\in \CIc(\RR^n\setminus\overline{\OO})$ such that
\begin{equation}
\label{eqn:u}
    u := (P_{\Gamma}-z_0)^{q_0 -1} \Pi_{\Gamma}(\Omega)w \neq 0,\quad\textrm{here } z_0=z(\id). 
\end{equation}    
Fixing $x_0\in\partial\OO$ and $h\in(0,h_0]$, we define $\varphi_h^{t}$ as in Step 5 and write $\Phi_t:=\varphi_h^{t}$, $t\in(-t_0,t_0)$. If we set \begin{equation}
\label{eqn:u_t}
        u(t) := ({\Phi_t}^{-1})^* v(t),\  v(t) := (P_{\Gamma,\Phi_t}-z(t))^{q_0 -1} \Pi_{\Gamma,\Phi_t}(\Omega) w,\  z(t):=z(\Phi_t).
\end{equation} 
Then by \eqref{bad case 2} we have for any 
\begin{equation}
\label{bad case 2 for t}
    \forall\,t\in(-t_0,t_0),\quad
    m_{\Gamma,\Phi_t}(z(t)) = \rank \Pi_{\Gamma,\Phi_t}(\Omega) = M,\  (P_{\Gamma,\Phi_t}-z(t))v(t)=0.
\end{equation}
Recalling \eqref{eqn:P1 Phi} and \eqref{eqn:P1Phi ext PGammaPhi}, we obtain the equation for $u(t)$:
\begin{equation}
\label{eqn:outgoing u_t}
    (-\Delta-z(t))u(t) = 0 \quad\textrm{on }\RR^n\setminus\Phi_t(\OO),
\end{equation}
in the sense of $L^2_{\loc}$ functions.

We next aim to show that $z(t)$ is differentiable at $0$. For that we extend \eqref{phi h deformation} to $\Phi_t\in\CI(\RR^n,\CC^n)$, $t\in\CC$:
\[
    \Phi_t(x) := x + t V_h(x),\quad t\in\CC
\]
We set $t_1=t_1(h)$ sufficiently small such that for all $|t|<t_1$ and $x\in\RR^n$, $D\Phi_t(x)= I + t D V_h(x)$ is invertible. Denoting by $ J_{\Phi_t} ^{ij} (x) = [D\Phi_t(x)^{-1}]_{ij}$, $\Phi_t^m(x)$ the $m$-th component of $\Phi_t(x)$, we replace $\Phi$ by $\Phi_t$ in \eqref{eqn:V} to define
\[
    \mathcal{V}(t) := \sum_{i,j} a_{i j}(t,x)\partial_{x_i x_j}^2 + \sum_j b_j(t,x)\partial_{x_j},
\]
\[
    \textrm{where}\quad a_{i j} = \delta_{i j} - \sum_\ell J_{\Phi_t}^{i \ell} J_{\Phi_t}^{j\ell},\quad  b_j =  - \sum_{i \ell m q} (\partial_{x_i x_\ell}^2 \Phi_t^m) J_{\Phi_t}^{j m} J_{\Phi_t}^{iq} J_{\Phi_t}^{\ell q}.
\]
Repeating the calculation that yields \eqref{eqn:coeffs V}, we also have for some $C=C(h)>0$,
\[
    a_{ij}(t,\cdot),\ b_j(t,\cdot) \in \CIc(\RR_x^n),\quad \|a_{ij}(t,\cdot)\|_\infty,\, \|b_j(t,\cdot)\|_\infty < C |t|.
\]
It follows that $\mathcal{V}(t)$, $|t|<t_1$ satisfies Hypothesis \ref{hypo:analytic perturbation}. For $t\in \CC$, $|t|<t_1$, we follow \eqref{eqn:PGamma t} to define
\[
     P_\Gamma(t) = P_\Gamma + \mathcal{V}(t),\quad\textrm{with domain }\D(P_\Gamma(t)):= \D(P_\Gamma).
\]
Recalling Propositions \ref{prop:kato perturbation} and \ref{prop:analytic dependence}, decreasing $t_0$ if necessary, for any $t\in\CC$, $|t|<t_0$, $P_\Gamma(t)$ has a discrete spectrum in some neighborhood $K$ of $z_0$ and the total multiplicity of the eigenvalues of $P_\Gamma(t)$ in $K$ equals $M$. Moreover, if we denote by $z_1(t),\ldots,z_M(t)$ the eigenvalues of $P_\Gamma(t)$ in $K$, repeated with multiplicity, then $\hat{z}(t) = M^{-1} \sum_{j=1}^M z_j(t)$ is an analytic function in $t\in\CC$, $|t|<t_0$. On the other hand, if we consider real $t$, $t\in(-t_0,t_0)$, then \eqref{eqn:V} and \eqref{eqn:P_Gamma,Phi} imply that 
\[
   P_\Gamma(t) = P_\Gamma + \mathcal{V}(t) = P_{\Gamma,\Phi_t},\quad -t_0<t<t_0.
\]
It follows from \eqref{bad case 2 for t} that for $t\in(-t_0,t_0)$ the eigenvalues of $P_\Gamma(t)$ near $z_0$ don't split, i.e. $z_j(t) = z(t)$, $j=1,\ldots,M$. Thus $z(t)=\hat{z}(t)$ when $t$ is real, $t\in(-t_0,t_0)$. The analyticity of $\hat{z}(t)$ gives the smoothness of $z(t)$ on $(-t_0,t_0)$. As a consequence, $u(t)$ and $v(t)$ defined in \eqref{eqn:u_t} also depend smoothly on $t\in(-t_0,t_0)$.

Since $\Phi_t(\OO)\subset\OO$ for $t\geq 0$, we can restrict \eqref{eqn:outgoing u_t} to the region $\RR^n\setminus\OO$ then differentiate the equation in $t$, by taking $t=0$, we obtain that
\begin{equation}
\label{eqn:u'(0)}
    (-\Delta-z_0) \partial_t u(0,x) = z'(0) u(x) \quad\textrm{on }\RR^n\setminus\OO.
\end{equation}
We recall \eqref{eqn:u_t} that $u(t,x) = v(t,\Phi_t^{-1}(x))$, using $u(0,x)=v(0,x)=u(x)$ and \eqref{phi h deformation} we can calculate the derivative in $t$:
\[ \partial_t u(0,x) = \partial_t v(t,\Phi_t^{-1}(x))|_{t=0} = \partial_t v(0,x) - \partial_x u \cdot V_h(x). \]
In view of \eqref{eqn:EGamma} and \eqref{eqn:u}, $u\in\mathcal{E}_\Gamma(z_0)$ is a resonant state of $-\Delta_\OO$ at $z_0$, thus we recall \cite[Theorem 4.7]{res} that $u\in\CI(\RR^n\setminus\OO)$. Then by \eqref{eqn:Vh} we conclude that
\begin{equation}
\label{eqn:f}
\begin{gathered}
    (-\Delta - z_0) (\partial_t v(0,x) - f) = z'(0)u(x) \quad\textrm{on }\RR^n\setminus\OO, \\
    f:=\partial_x u\cdot V_h(x)\in\CIc(\RR^n\setminus\OO),\quad f|_{\partial\OO} = \chi_h  \partial_\nu u.
\end{gathered}
\end{equation}
It follows from $v(t,x)\in\D(P_\Gamma),\ t\in(-t_0,t_0)$ that $\partial_t v(0,x)\in D(P_\Gamma)$, thus the first equation in \eqref{eqn:f} reduces to
\begin{equation}
\label{final equation}
    (P_\Gamma -z_0)\partial_t v(0,x) = (-\Delta-z_0)f + z'(0)u \quad\textrm{on }\RR^n\setminus\OO.
\end{equation}
We introduce the bilinear form on $B_0\times B_1$ (no complex conjugation),
\[ \langle u, v\rangle := \int_{\RR^n\setminus\OO} uv\, dx,\quad u\in B_1,\ v\in B_0.\]
We now apply the projection $\Pi_\Gamma$ (omitting $\Omega$) to both sides of \eqref{final equation}, pair with $(P_\Gamma - z_0)^{q_0-1} w\in B_0$ (since $w\in\CIc(\RR^n\setminus\overline{\OO})$), use the fact that $(P_\Gamma - z_0)\Pi_\Gamma g = \Pi_\Gamma (P_\Gamma - z_0) g$, $\forall\,g\in\D(P_\Gamma)$ to obtain that
\[ \begin{split}
    &{\ }\quad \langle (P_\Gamma - z_0)\Pi_\Gamma \partial_t v(0,x) , (P_\Gamma -z_0)^{q_0-1} w \rangle \\
    &=  \langle \Pi_\Gamma(-\Delta-z_0)f, (P_\Gamma -z_0)^{q_0-1} w \rangle 
     + z'(0) \langle u, (P_\Gamma - z_0)^{q_0-1}w\rangle.
\end{split}
\]
By Green's formula, $\langle P_\Gamma g_1, g_2\rangle = \langle g_1, P_\Gamma g_2\rangle$ for any $g_1\in \D(P_\Gamma)$, $g_2\in\CIc(\RR^n\setminus\overline{\OO})$. It then follows from \eqref{bad case 2} and \eqref{eqn:u} that
\[
\langle (P_\Gamma - z_0)\Pi_\Gamma \partial_t v(0,x) , (P_\Gamma -z_0)^{q_0-1} w \rangle = \langle (P_\Gamma-z_0)^{q_0} \Pi_\Gamma \partial_t v(0,x), w\rangle=0,
\]
and that 
\[
\langle u,(P_\Gamma -z_0)^{q_0 - 1}w\rangle = \langle (P_\Gamma -z_0)u,(P_\Gamma -z_0)^{q_0 - 2}w\rangle = 0.
\]
Since $\langle\Pi_\Gamma f_1,f_2\rangle = \langle \Pi_\Gamma f_2, f_1\rangle$ for any $f_1,f_2\in B_0$, we conclude that
\[
\begin{split}
    0 & = \langle \Pi_\Gamma(-\Delta-z_0)f, (P_\Gamma -z_0)^{q_0-1} w \rangle \\
     & = \langle (-\Delta-z_0)f,(P_\Gamma -z_0)^{q_0-1} \Pi_\Gamma w \rangle = \langle (-\Delta-z_0) f , u \rangle.
\end{split}
\]
Now we apply Green's formula and recall \eqref{eqn:f} and $u_{\partial\OO}=0$ to obtain
\[ 
0 = \langle (-\Delta-z_0) f , u \rangle = \int_{\partial\OO} f\,\partial_\nu u \,dS = \int_{\partial\OO} \chi_h(x') (\partial_\nu u(x'))^2 dS(x') .
\]
Since the above equation holds for any $h\in(0,h_0]$, ($u$ is independent of $x_0$ and $h$) sending $h$ to $0+$, by \eqref{eqn:chi_h} we can derive that $\partial_\nu u(x_0) = 0$. We note that $x_0\in\partial\OO$ can be chosen arbitrarily, thus $\partial_\nu u|_{\partial\OO} \equiv 0$. However, it follows from \eqref{bad case 2} and \eqref{eqn:u} that $u\in\D_1(\OO)$ satisfying $(-\Delta-z_0)u=0$ on $\RR^n\setminus\OO$. Extending $u$ into $\OO$ by $u|_\OO=0$, it then follows from \eqref{eqn:outgoing u_t} and the boundary values $u|_{\partial\OO} = 0$, $\partial_\nu u|_{\partial\OO}=0$ that $u\in H_{\loc}^1(\RR^n)$ is a weak solution of $(-\Delta-z_0)u=0$ on $\RR^n$. The unique continuation property of second order elliptic differential equations shows that $u\equiv 0$, which contradicts \eqref{eqn:u}.

\noindent
Step 7. It remains to consider the case $q_0=1$ in \eqref{bad case 2}. Let $\{w_j\}_{j=1}^M$ be a set of vectors in $\CIc(\RR^n\setminus\overline{\OO})$ such that $\{\Pi_\Gamma w_j\}_{j=1}^M$ is a basis for $\Ran\Pi_\Gamma$. Since $\Pi_\Gamma$ is symmetric with respect to the bilinear form $\langle\cdot,\cdot\rangle$ on $B_0\times B_0$, the matrix $A$, $A_{ij}:=\langle \Pi_\Gamma w_i,w_j \rangle$ is a complex symmetric matrix. To see $A$ is nondegenerate, we suppose that 
\[ \exists\,x\in\CC^M,\quad \langle \Pi_\Gamma w_i,\sum_j x_j w_j\rangle = 0,\ i=1,\cdots,M.\]
Since $\{\Pi_\Gamma w_i\}_{i=1}^M$ spans $\Ran\Pi_\Gamma$, we have $\langle \Pi_\Gamma w , \sum x_j w_j\rangle = 0$ for all $w\in B_0$, which implies that $\langle \sum x_j\Pi_\Gamma w_j,w\rangle = 0$, $\forall\,w\in B_0$. Hence $\sum x_j\Pi_\Gamma w_j = 0\Rightarrow x=0$. We apply the Takagi factorization to the matrix A to obtain that
\[ A = U^T \diag(r_1,\cdots,r_M) U,\ \textrm{where $U$ is unitary, $r_j^2$ are the eigenvalues of $A A^* $.  }  \]
We remark that $U^T$ is the real transpose. Then we can write $A = B^T B$, $B$ nondegenerate due to the nondegeneracy of $A$. Transforming $\{w_j\}_{j=1}^M$ by the matrix $B$ and putting $u_j:=\Pi_\Gamma w_j$, we may assume now that    
\[ \Ran\Pi_\Gamma = \Span\{u_j\}_{j=1}^M,\quad \langle u_j, w_i\rangle = \delta_{ij}.   \]
For any fixed $x_0\in\partial\OO$ and $h\in(0,h_0]$, we define the evolution of each $u_j$ as in \eqref{eqn:u_t}:
\begin{equation}
\label{eqn:uj_t}
    u_j(t):=(\Phi_t^{-1})^* v_j(t),\quad v_j(t):=\Pi_{\Gamma,\Phi_t}(\Omega)w_j,\ z(t):=z(\Phi_t).
\end{equation}
We note that \eqref{final equation} still holds with $\partial_t v(0,x),u,f$ replaced by $\partial_t v_j(0,x),u_j$ and $f_j$ defined as in \eqref{eqn:f}. The same arguments as in Step 6 show that  
\[ \langle (P_\Gamma -z_0)\Pi_\Gamma v_j'(0),w_i\rangle = \langle \Pi_\Gamma(-\Delta-z_0) f_j,w_i\rangle + z'(0)\langle u_j,w_i\rangle. \]
Since $(P_\Gamma -z_0)\Pi_\Gamma = 0$ by \eqref{bad case 2} with $q_0=1$, it then follows that
\[  \langle (-\Delta-z_0)f_j , u_i \rangle = -z'(0)\delta_{ij}. \]
We apply Green's formula with boundary value of $f_j$ like \eqref{eqn:f} to obtain that  
\[ 
- z'(0)\delta_{ij} = \langle (-\Delta-z_0)u_i,f_j\rangle + \int_{\partial\OO} f_j\partial_\nu u_i \,dS = \int_{\partial\OO}\chi_h (\partial_\nu u_i)(\partial_\nu  u_j) \,dS.
\]
Since $M\geq 2$, for any $x_0\in\partial\OO$ and $h\in(0,h_0]$ we have
\[ \int_{\partial\OO} \chi_h (\partial_\nu u_1)^2 dS = \int_{\partial\OO} \chi_h (\partial_\nu u_2)^2 dS,\quad \int_{\partial\OO} \chi_h \partial_\nu u_1 \partial_\nu u_2\,dS = 0. \]
Sending $h\to 0+$, it follows from \eqref{eqn:chi_h} that
\[ (\partial_\nu u_1(x_0))^2 = (\partial_\nu u_2(x_0))^2,\quad \partial_\nu u_1 (x_0) \partial_\nu u_2 (x_0) = 0, \]
thus $\partial_\nu u_1 (x_0) = \partial_\nu u_2 (x_0) = 0$. Since $x_0\in\partial\OO$ i arbitrary, $\partial_\nu u_1\equiv 0$. Hence the same arguments as in the end of Step 6 show that $u_1\equiv 0$, which gives a contradiction.
\end{proof}

\def\arXiv#1{\href{http://arxiv.org/abs/#1}{arXiv:#1}}

\end{document}